\documentclass[acmtog]{acmart}

\acmJournal{TOG}

\newcommand*{\ShowNotes}{}
\usepackage[norelsize, algoruled, vlined, noend]{algorithm2e}

\definecolor{darkred}{rgb}{0.7,0.1,0.1}
\definecolor{darkgreen}{rgb}{0.1,0.5,0.1}
\definecolor{cyan}{rgb}{0.7,0.0,0.7}
\definecolor{dblue}{rgb}{0.2,0.2,0.8}
\definecolor{maroon}{rgb}{0.76,.13,.28}
\definecolor{burntorange}{rgb}{0.81,.33,0}
\definecolor{royalpurple}{rgb}{0.47,.31,0.66}

\ifdefined\ShowNotes
  \newcommand{\colornote}[3]{{\color{#1}\bf{#2 #3}\normalfont}}
\else
  \newcommand{\colornote}[3]{}
\fi

\definecolor{mygray}{gray}{0.6}
\definecolor{keywordcolor}{rgb}{0.2,0.2,0.6}
\definecolor{keywordcolor2}{rgb}{0.15,0.46,0.1}
\definecolor{typecolor}{rgb}{0.17,0.56,0.68}
\definecolor{commentcolor}{gray}{0.3}
\definecolor{ratecolor}{rgb}{0.5,0.1,0.1}
\definecolor{stringcolor}{gray}{0.3}

\newcommand\todosilent[1]{}

\begin{document}

\title{Vid2Player: Controllable Video Sprites that Behave and Appear like Professional Tennis Players}

 \author{Haotian Zhang}
 \email{haotianz@stanford.edu}
 \affiliation{\institution{Stanford University}}
 
 \author{Cristobal Sciutto}
 \email{csciutto@stanford.edu}
 \affiliation{\institution{Stanford University}}
 
 \author{Maneesh Agrawala}
 \email{maneesh@cs.stanford.edu}
 \affiliation{\institution{Stanford University}}
 
 \author{Kayvon Fatahalian}
 \email{kayvonf@cs.stanford.edu}
 \affiliation{\institution{Stanford University}}

\renewcommand{\shortauthors}{Zhang and Sciutto et al.}


\begin{abstract}

We present a system that converts annotated broadcast video of tennis matches into interactively controllable video sprites that behave and appear like professional tennis players.
Our approach is based on controllable video textures, and utilizes domain knowledge of the cyclic structure of tennis rallies to place clip transitions and accept control inputs at key decision-making moments of point play.
Most importantly, we use points from the video collection to model a player's court positioning and shot selection decisions during points.
We use these behavioral models to select video clips that reflect actions the real-life player is likely to take in a given match play situation,
yielding sprites that behave realistically at the macro level of full points, not just individual tennis motions.
Our system can generate novel points between professional tennis players that resemble Wimbledon broadcasts,
enabling new experiences such as the creation of matchups between players that have not competed in real life,
or interactive control of players in the Wimbledon final.
According to expert tennis players, the rallies generated using our approach are significantly more realistic in terms of player behavior than video sprite methods that only consider the quality of motion transitions during video synthesis.


\end{abstract}



\begin{teaserfigure}
\includegraphics[width=\textwidth]{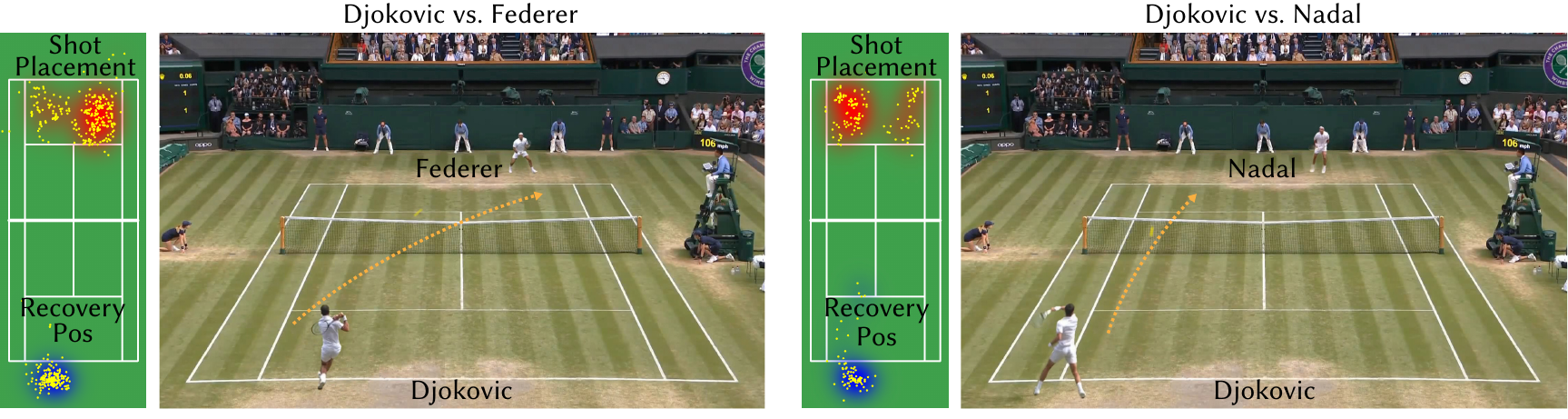}
\caption{
Our system converts annotated broadcast video of Wimbledon tennis matches
into controllable video sprites that appear and behave like professional tennis players.
Our system uses player and ball trajectories extracted from video
to construct behavior models that predict a player's shot placement
(where they hit the ball to) and where they position themself on the court against a specific opponent.
These models provide control inputs to a video-textures system that leverages domain knowledge of the cyclic struture of tennis rallies to make clip transitions at key-decision making moments of play.
As a result, our video sprites play points with similar style and strategy as their real-world counterparts.
Visualizations depict where our model of Novak Djokovic hits the ball to (shot placement)
and positions himself (recovery position) when hitting backhands against Roger Federer and Rafael Nadal.
Djokovic's strategy changes to hit to the right side of the court vs. Federer (his weaker backhand side)
while a majority of shots vs. Nadal go to the left.  
}
\label{fig:teaser}
\end{teaserfigure}

\maketitle

\section{Introduction}
\label{sec:intro}

From high-school games to professional leagues,
it is common for athletic competitions 
to be captured on high-resolution video.
These widely available videos provide a rich source of information about 
how athletes look (appearance, style of motion)
and play their sport (tendencies, strengths, and weaknesses).
In this paper we take inspiration from the field of sports analytics,
which uses analysis of sports videos to create predictive models of athlete behavior.
However, rather than use these models to inform coaching tactics or improve play,  
we combine sports behavioral modeling with image-based rendering 
to construct interactively controllable video sprites that both
\emph{behave and appear} like professional tennis athletes.

Our approach takes as input a database of broadcast tennis videos,
annotated with important match play events
(e.g., time and location of ball contact, type of stroke).
The database contains a few thousand shots from each player,
which we use to construct behavior models of how the player positions themselves on the court
and where they are likely to hit the ball in a given match play situation.
These behavioral models provide control inputs to 
an image-based animation system based on controllable
video textures\,\cite{Schodl:2000:videotextures,Schodl:2002:spriteanim}.
All together, the system interactively generates realistic 2D sprites of
star tennis players playing competitive points.

Our fundamental challenge is to create a video sprite of a tennis player
that realistically reacts to a wide range of competitive point play situations,
given videos for only a few hundred points.  
To generate realistic output from limited inputs,
we employ domain knowledge of tennis strategy and point play 
to constrain and regularize data-driven techniques for player behavior modeling and video texture rendering.
Specifically, we make the following contributions:

\textbf{Shot-cycle state machine.}
We leverage domain knowledge of the structure of tennis rallies
to control and synthesize video sprites at the granularity of \emph{shot cycles}---a
sequence of actions a player takes to prepare, hit, and recover from
each shot during a point.
Constructing video sprites at shot-cycle granularity has multiple benefits. 
It reduces the cost of searching for source video clips in the database
(enabling interactive synthesis),
constrains video clip transitions to moments where fast-twitch
player motion makes transition errors harder to perceive (improving visual quality),
and aligns the control input for video sprite synthesis
with key decision-making moments during point play.

\textbf{Player-specific, data-driven behavior models.}
We build models that predict a player's court positioning and
shot selection, which are used as control inputs during video sprite synthesis.
By combining both visual-quality metrics and the player-specific behavior models to guide video synthesis,
we create video sprites that are ``realistic'' both at a fine-grained level in that they depict a
player's actual appearance and style of motion,
but also at a macro level in that they capture real-life strategies (hitting the ball to an opponent's weakness) and tendencies (aggressive vs. defensive court positioning) during point play.

\textbf{High-quality video sprite rendering on diverse, real-world data.}
We provide methods for preparing a large database of real-world,
single-viewpoint broadcast video clips for use in a controllable video sprites system.
This involves use of neural image-to-image transfer methods to eliminate 
differences in player appearance over different match days and times,
hallucinating missing pixel data when players are partially cropped in the frame,
and making video sprite rendering robust to errors 
in annotations generated by computer vision techniques (e.g. player detection, pose estimation).

We demonstrate that our system can synthesize videos 
of novel points that did not occur in real life,
but appear as plausible points between star athletes on Wimbledon's Centre Court (Fig.~\ref{fig:teaser}).
This also enables new experiences such as creating points between players that have never competed in real life
(such as Roger Federer and Serena Williams),
or interactive experiences where a user
can control Roger Federer as he competes against a rival in the Wimbledon final.
An evaluation with expert tennis players shows that the rallies generated using our approach are significantly more realistic in terms of player behavior than video sprite methods that only consider 
video clip transition quality and ball contact constraints during video synthesis.


\section{Related Work}
\label{sec:related}

\paragraph{Controllable video textures.}
We adopt an image-based rendering approach that assembles frames for a novel animation from a database of broadcast video clips\,\cite{Schodl:2000:videotextures,Schodl:2002:spriteanim,Efros:2003:actiondistance,Flagg:2009:humanvideotex}.  
Most prior work on generating controllable video sprites focuses on finding clip transitions
where the local frame-to-frame motion and appearance are similar.
While results are plausible frame-to-frame,
higher-level behavior of the character can be unrealistic.
Our player behavior models provide higher level control inputs that yield 
player sprites that behave realistically over the span of entire points.

The Tennis Real Play system\,\cite{Lai:2012:TennisRealPlay} also creates controllable video sprite based tennis characters from broadcast tennis videos and structures synthesis using a simple \emph{move-hit-standby} state machine that is similar to our shot cycle state machine described in Section~\ref{sec:rallystructure}
(they also use match play statistics to guide shot selection to a limited extent). 
However, this prior work does not generate realistic player motion or behavior
(see supplementary video for example output from Tennis Real Play).
We utilize a video database that is nearly two orders of magnitude larger,
a shot cycle state machine that differentiates pre-shot and post-shot movement,
and more complex player behavior models
to generate output that appears like the players as seen in broadcast video footage.

\paragraph{Motion graphs.}
Using video analysis to extract player trajectory and pose information from source video\,\cite{Girdhar:2018:detecttrack,Guler:2018:densepose} reduces the problem of finding good video transitions to one of finding a human skeletal animation that meets specified motion constraints.  
Rather than stitch together small snippets of animation to meet motion goals
(e.g., using expensive optimization techniques for motion planning or motion graph search)\,\cite{Kovar:2002:motiongraphs,Lee:2002:interactivecontrol},
we adopt a racket-sports-centric organization of a large video database,
where every clip corresponds to a full shot cycle of player motion.
This allows us to limit transitions to the start and end of clips,
and allows a large database to be searched in real time for good motion matches.

\paragraph{Conditional Neural Models.}
Vid2Game\,\cite{gafni:2019:vid2game} utilizes end-to-end learning to create models that 
animate a tennis player sprite in response to motion control inputs from the user.
However Vid2Game does not generate video sprites that move or play tennis in a realistic manner:
player swings do not contact an incoming ball and player movement is jerky (see supplemental video for examples). 
While end-to-end learning of how to play tennis is a compelling goal,
the quality of our results serves as an example of the benefits of injecting domain knowledge 
(in the form of the shot cycle structure and a data-driven behavioral model) into learned models.

\paragraph{Image-to-image transfer.}
Our system utilizes both paired\,\cite{Isola:2017:pix2pix, Wang:2018:pix2pixHD} and unpaired\,\cite{Zhu:2017:cycleGAN}) neural image-to-image transfer to normalize lighting conditions and player appearance (e.g., different clothing) across a diverse set of real-world video clips.
Many recent uses of neural image transfer in computer graphics are paired transfer scenarios that enhance the photorealism of synthetic renderings\,\cite{wang:2018:vid2vid,kim:2018:deepvideoportrait,bi:2019:deepcg2Real,Chan:2019:everybodydance,Liu:2019:body2Body}. 
We use this technique for hallucinating occluded regions of our video sprites.
Like prior work transforming realistic images from one visual domain to another\,\cite{choi:2018:stargan,chang:2018:pairedcyclegan},
our system uses unpaired transfer for normalizing appearance differences in lighting conditions and clothing in our video sprites. 

\paragraph{Play forecasting for sports analytics.}
The ability to capture accurate player and ball tracking data in professional sports venues\,\cite{StatsPerformSportVU,SecondSpectrum,Owens:2003:hawkEye} has spurred interest in aiding coaches and analysts with data-driven predictive models of player or team behavior.
Recent work has used multiple years of match data to predict where tennis players will hit the ball\,\cite{Wei:2016:shotforecasting,Fernando:2019:forecasting},
whether a point will be won\,\cite{Wei:2016:thinedge},
how NBA defenses will react to different offensive plays\,\cite{Le:2017:ghosting},
and analyze the risk-reward of passes in soccer\,\cite{Power:2017:soccerpasses}.
We are inspired by these efforts, and view our work as connecting the growing field of sports play forecasting with the computer graphics challenges of creating controllable and visually realistic characters.

\section{Background: Structure of Tennis Rallies}
\label{sec:rallystructure}

\begin{figure}[t]
	\centering
	\includegraphics[width=\columnwidth]{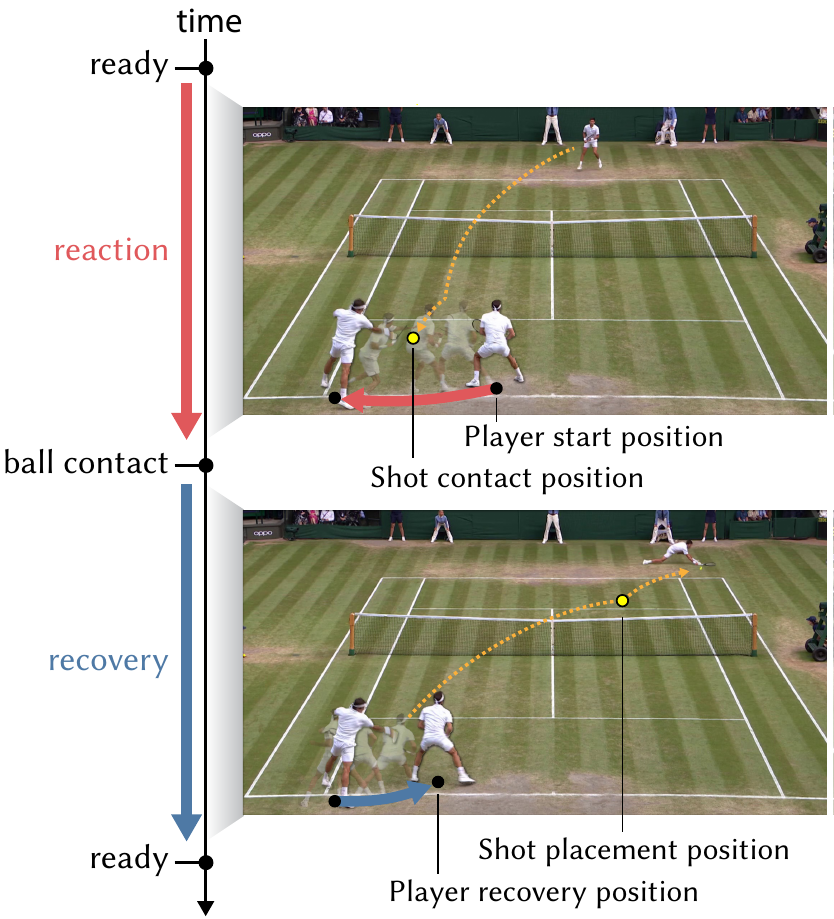}
	\caption{
		A shot cycle for the player on the near side of the court.  The cycle begins with the player in the ``ready'' state just as the opponent strikes the ball.  Next, the player moves to hit the ball (``reaction'' phase). He makes contact, striking the ball so that it lands on his opponent's side of the court. After contact, the player returns to the ready position in a new court location to prepare for his opponent's next shot (``recovery'' phase).}
    \label{fig:movement}
\end{figure}

Many aspects of our system's design rely on domain knowledge of the structure of tennis rallies\,\cite{smith:2017:absolutetennis}.
As background for the reader, we describe key elements of a tennis rally here.

In tennis, play is broken into points,
which consist of a sequence of \emph{shots} (events where a player strikes the ball).  
A point begins with a serve and continues with a \emph{rally} where opposing players alternate hitting shots that return the ball to their opponent's side of the court.
The point ends when a player fails to successfully return the incoming ball, either because they can not reach the ball (the opponent hits a \emph{winner}) or because the opponent hits the ball into the net or outside the court (the opponent makes an \emph{error}).

During a rally, players make a critical sequence of movements and decisions in the time between two consecutive shots by their opponent.
We refer to this sequence, which involves reacting to the incoming ball, hitting a shot, and finally recovering to a position on the court that anticipates the next shot, as a \emph{shot cycle}. Fig.~\ref{fig:movement} illustrates one shot cycle for the player on the near side of the court.

\emph{Phase 1: (Ready$\rightarrow$Reaction$\rightarrow$Contact).}
At the start of the shot cycle
(when the player on far side of the court in Fig.~\ref{fig:movement} strikes the ball)
the near player is posed in a stance that facilitates a quick reaction to the direction of the incoming shot.
We mark this moment as \emph{ready} on the timeline.
Next, the near player reacts to the incoming ball by maneuvering himself to hit a shot.
It is at this time that the player also makes \emph{shot selection} decisions, which in this paper include:
\emph{shot type}, what type of shot to use (e.g., forehand or backhand, topspin or underspin); \emph{shot velocity}, how fast to hit the shot; and \emph{shot placement position},
where on the opponent's side of the court they will direct their shot to.
We refer to the period between ready and the time of \emph{ball contact} as the \emph{reaction} phase of the shot cycle. 

\emph{Phase 2: (Contact$\rightarrow$Recovery$\rightarrow$Ready).} After making contact and finishing the swing, the near player moves to reposition themself on the court in anticipation of the opponent's next shot.
This movement returns the player back to a new ready position right as the ball reaches the opponent.
We call this new ready position the \emph{recovery position} and the time between the player's ball contact and arrival at the recovery position the \emph{recovery} phase. 

The shot cycle forms a simple state machine that governs each player's behavior during a rally.
In a rally, both players repeatedly execute new shot cycles,
with the start of the reaction phase of one player's shot cycle coinciding with the start of the recovery phase in the current shot cycle of their opponent. (The start of a shot cycle for one player is offset from that of their opponent by half a cycle.)
This state machine forms the basis of our approach to modeling player behavior and generating video sprites.

\section{System Overview}
\label{sec:overview}

\begin{figure*}[t]
	\centering
	\includegraphics[width=\textwidth]{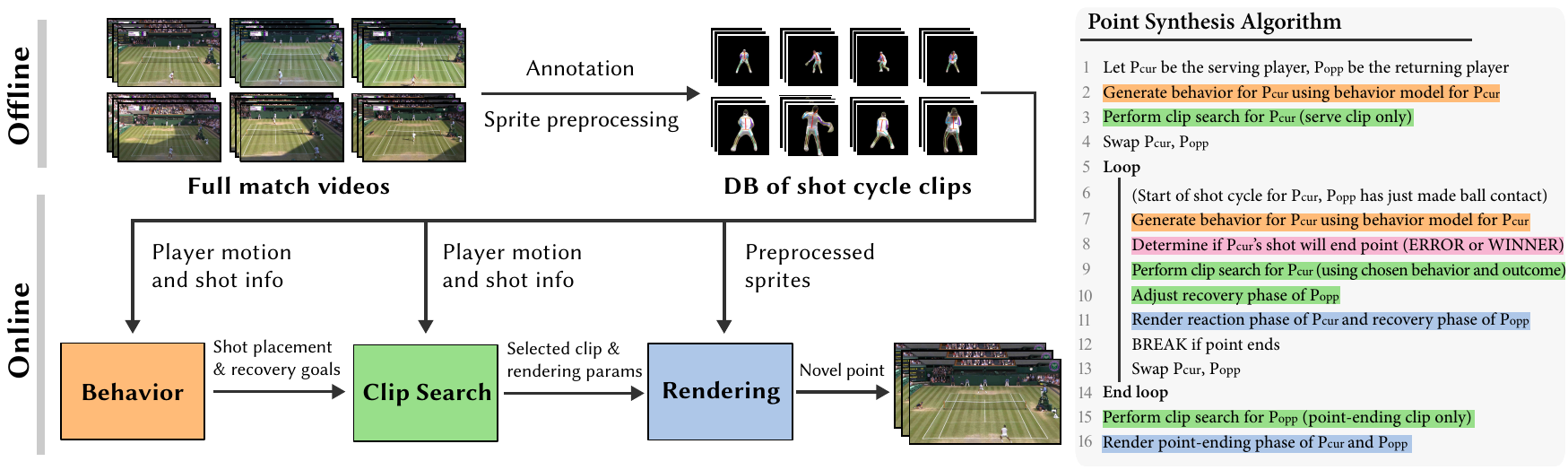}
	\caption{System overview. Left. Offline preprocessing prepares a database of annotated shot cycle clips.  The annotated clips are used as inputs for modeling player behavior, choosing video sprites that meet behavior goals, and sprite-based rendering.  Right: Pseudocode for generating a novel point.
	Each iteration of the loop corresponds to the start of a new shot cycle for one of the two players.
	We color system component boxes and lines of code to show correspondence.}
	\label{fig:overview}
\end{figure*}

Fig.~\ref{fig:overview} summarizes our point synthesis system.
During an offline preprocessing phase,
we annotate input match videos to create a database of clips,
each depicting one shot cycle of play
(Sec.~\ref{sec:dataset}).
Player motion and shot information extracted from shot cycle clips
serve as training data for behavior models that
determine player goals during a point (Sec.~\ref{sec:behavior}).
We also preprocess database clips to identify incomplete (cropped) player sprites 
and correct for appearance differences that would create discontinuous video texture output
(Sec.~\ref{sec:rendering_preprocess}).

Given the database of annotated shot cycles and player behavior models for
two players (one server, one returner),
the system synthesizes video of a novel point, shot cycle by shot cycle,
as given by the pseudocode (Fig.~\ref{fig:overview}-right).
The algorithm begins by choosing a serving clip to start the point.  
Then, each iteration of the main loop corresponds to the start of a new shot cycle for one of the two players.
For example, the first iteration corresponds to the start of the serve returner's first shot cycle.
In each iteration, the system invokes the player's behavior model to generate goals for the player in response to the incoming ball (e.g., what type of shot to hit, where to place the shot, where to recover to on the court).
Then the system performs a database search for a shot cycle clip that meets the player's behavior goals (Sec.~\ref{sec:motion}),
Finally the system renders video frames for both players
by compositing video sprites for the chosen clips into a virtual court environment (Sec.~\ref{sec:rendering}).  Rendering proceeds up until the point of ball contact
(the reaction phase of the shot cycle for the current player, and the recovery phase for the opponent.)
The process then repeats again to start a new shot cycle for the opponent.
In each iteration of the loop, the system also determines if the point should end,
either because the player's shot was an error or was unreachable by the opponent.
Although not shown in the pseudocode,
the algorithm can also receive player behavior goals directly from interactive user input (Section~\ref{sec:eval:interactive}).

Constructing player video sprites at the coarse granularity of shot cycles yields several benefits.
First, making all control decisions at the start of the shot cycle aligns with the moment when real-life tennis players make key decisions about play.  This allows clip selection to consider goals for all phases of the shot cycle: shot preparation (reaction), swing, and recovery to identify a continuous segment of database video that globally meet these goals.  Second, placing transitions at the end of the shot cycle as the player nears the ready position makes discontinuities harder to perceive, since this is the moment where tennis players make abrupt changes in motion as they identify and react to the direction of an incoming shot.
It is also the moment when the opposing player is hitting the ball,  
so the viewer's gaze is likely to be directed at the ball rather than at the player undergoing a clip transition. 
Finally, performing clip transitions at shot-cycle granularity facilitates interactive performance,
since search is performed infrequently and only requires identification of single matching database clip
(not a full motion graph search).







\section{Tennis Video Database}
\label{sec:dataset}

We use videos of Wimbledon match play as source data for our system.
The dataset consists of three matches of Roger Federer, Rafael Nadal, Novak Djokovic playing against each other along with two matches of Serena Williams playing against Simona Halep and Camila Giorgi.
All matches took place on Wimbledon's Centre Court during the 2018 and 2019 Wimbledon tournaments.  We 
retain only video frames from the main broadcast camera, and discard all clips featuring instant replays or alternative camera angles.

\paragraph{Shot cycle boundaries and outcome.} We manually annotate the identity of the player on each side of the court, 
the video time of the beginning and ending of all points,
and the time when players hit the ball.
We use these as time boundaries to organize the source video into clips,
where each clip depicts a player carrying out one shot cycle.
(Note that each frame in the source video that lies during point play exists in two shot cycle clips, one for each player.) 
Table~\ref{tab:datasetstats} gives the number of clips in the database and
the total time spanned by these clips.
Table~\ref{tab:clipschema} describes the schema for a database clip.

For each clip $i$, we store the time of ball contact ($t_c^i$)
and time when the recovery phase ends ($t_r^i$).  
We identify the type of the shot hit during the shot cycle ($c^i$):
topspin groundstrokes, underspin groundstrokes,
volleys, and serves\,\cite{smith:2017:absolutetennis}
(differentiating forehands and backhands for groundstrokes and volleys).
We also label the outcome of the shot cycle ($o^i$):
the player hits the ball in the court without ending the point,
hits a winner, hits an error, or does not reach the incoming ball (no contact).

\begin{table}[t]
	\resizebox{\linewidth}{!}{
		\centering
		\begin{tabular}{l r r r r r r r r r}
			& S   & FH-T & FH-U & BH-T & BH-U & FH-V & BH-V & Total & Dur (min)\\
			\hline
			R. Federer  & 241 & 576  & 31   & 343  & 240  & 24   & 29   & 1484  & 57.7 \\
			\hline
			R. Nadal    & 261 & 600  & 36   & 417  & 130  & 10   & 16   & 1470  & 56.9 \\
			\hline
			N. Djokovic & 340 & 759  & 49   & 744  & 100  & 7    & 9    & 2008  & 78.7 \\
			\hline
			S. Williams & 61  & 154  & 3    & 143  & 10   & 6    & 3    & 380   & 15.0 \\
			\hline
		\end{tabular}
	}
	
	\caption{
		Shot statistics in our tennis database.
		Shot types include: serve (S), topspin forehand/backhand (FH-T/BH-T), underspin forehand/backhand (FH-U/BH-U), and forehand/backhand volley (FH-V, BH-V).
		The most common shots are forehand and backhand topspin ground strokes.}
	\label{tab:datasetstats}
\end{table}

\begin{table}[t]
	\begin{tabular}{lp{2.85in}}
		\multicolumn{2}{l}{\textbf{Schema for video database clip $i$:}} \\
		\hline
		$t^i_{c}$                       &  Time of player's ball contact (length of reaction phase) \\
		$t^i_{r}$                  		&  Time of end of recovery phase (length of the clip)\\ 
		$c^i$							&  Shot type \\
		$o^i$							&  Shot outcome \\ 
		$\mathbf{x}^i_\mathbf{c}$       &  Shot contact position (court space) \\
		$t^i_{b}$                       &  Time of ball bounce on the court\\ 
		$\mathbf{x}^i_\mathbf{b}$       &  Shot placement position (court space) \\ [1em]
		
		\multicolumn{2}{l}{\vspace{.25em} For all frames in clip (indexed by time $t$):} \\
		$\mathbf{x}^i_\mathbf{p}(t)$    & Player's position (court space) \\
		$\mathbf{x}^i_\mathbf{o}(t)$    & Opponent's position (court space) \\
		$\mathbf{p}^i(t)$               & Player's pose (screen space, 14 keypoints) \\
		$\mathbf{b}^i(t)$               & Player's bounding box (screen space) \\
		$\mathbf{H}^i(t)$					& Homography mapping screen space to court space \\
		
		\hline
	\end{tabular}
	\caption{Each clip in the video database corresponds to a single shot cycle. All the time is relative to the start of the shot cycle and all the shot information is denoting the shot hit during this shot cycle. Annotation of the source video yields this per-clip metadata.
	}
	\label{tab:clipschema}
\end{table}

\paragraph{Player position and pose.}
For all frames in clip $i$ (indexed by time $t$), we estimate the screen space bounding box ($\mathbf{b}^i(t)$) and 2D pose ($\mathbf{p}^i(t)$) of the player undergoing the shot cycle,
and positions of both players on the ground plane in court space 
($\mathbf{x}^i_\mathbf{p}(t)$ and $\mathbf{x}^i_\mathbf{o}(t)$) using automated techniques.
We identify the player on the near side of the court using
detect-and-track\,\cite{Girdhar:2018:detecttrack}.
(We choose the track closest to the bottom of the video frame.)
Since detect-and-track occasionally confuses the track for the far court player with nearby linepersons,
to identify the far side player
we use Mask R-CNN\,\cite{He:2017:maskrcnn} to detect people above the net,
and select the lowest (in screen space) person that lies within three meters of the court.
For both players, we compute 2D pose and a screen space bounding box using Mask R-CNN.
Since the camera zooms and rotates 
throughout the broadcast, we use the method of Farin et al.\,\shortcite{farin:2003:courtlines} to detect lines on the court and estimate a homography relating the image plane to the court's ground plane ($\mathbf{H}^i(t)$).
We assume the player's feet are in contact with the ground,
and estimate the player's court space position by 
transforming the bottom center of their bounding
box into court space using this homography.

\begin{figure}[t]
	\centering
	\includegraphics[width=\columnwidth]{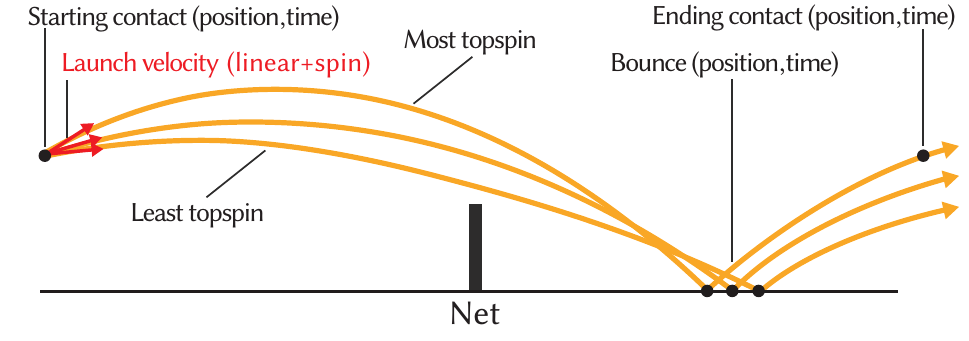}
	\caption{Given the space-time location (time, 3D position) of consecutive ball contacts during a rally, we find a ball trajectory that starts at the first space-time contact location, clears the net, bounces in the court, and then closely matches the ending location.  We use grid-search over the ball's initial linear velocity (horizontal and vertical components) and spin velocity to find a matching trajectory.}
	\label{fig:ball_trajectory}
\end{figure}

\paragraph{Ball trajectory.}

Tennis balls undergo significant motion blur in broadcast footage,
making it difficult to detect the ball from image pixels.
We take an alternative approach,
and estimate the ball's space-time trajectory given only the human-annotated ball contact times
and the position and pose of the swinging player during these times.
Specifically, we first estimate the 3D position of the ball at the time of 
contact (shot contact position $\mathbf{x}^i_\mathbf{c}$)
based on the court position and pose of the swinging player (details in supplemental material).
Then we estimate the ball's trajectory between two consecutive ball contacts in a rally by fitting a pair of ballistic trajectories (with a bounce in between) to the starting and ending contact points
(Fig.~\ref{fig:ball_trajectory}).

We adopt a physical model of ball flight:
\begin{align*}
	\frac{dv_x}{dt} &= -kv(C_d v_x + C_L v_y) \\
	\frac{dv_y}{dt} &= kv(C_L v_x - C_dv_y) - g
\end{align*}

Where $v$, $v_x$, and $v_y$ denote the magnitude of the ball's velocity, as well as its horizontal and vertical components. $k = \rho \pi R^2 / (2m)$ is a constant determined by the air density $\rho$, the ball's radius $R$ and mass $m$. 
$C_L$ is the lift coefficient due to Magnus forces created by the spin of the ball.
In tennis, topspin (ball rotating forward) imparts downward acceleration to the ball leading it to drop quickly.
Underspin (ball rotating backward) produces upward acceleration causing the ball to float\,\cite{tennisPhysicsBook:2004}.
$C_L$ is computed as $1/(2 + v/v_{\text{spin}})$ where $v_{\text{spin}}$ denotes the magnitude of ball's \emph{spin velocity}
(the relative speed of the surface of the ball compared to its center point).
We flip the sign of $C_L$ for topspin.
$C_d$ and $g$ refer to the air drag coefficient and magnitude of gravitational acceleration.
We simulate the ball's bounce as a collision where the coefficient of restitution is constant and the spin after bounce is always topspin\,\cite{tennisPhysicsBook:2004}. (We do not model the ball sliding along the court during a bounce.)
Given the 3D shot contact position and both the ball's linear and spin velocity at contact, we use the equations above to compute the ball's space-time trajectory before and after bounce.


Given the time and 3D positions of two consecutive ball contacts in a rally, 
we perform a grid search over the components of the ball's launch velocity (horizontal and vertical components of linear velocity, as well as spin velocity) to yield a trajectory which starts at the first contact point, clears the net, bounces in the court, and then closely matches the time and location of the second contact point.
This trajectory determines the ball's position at all times in between the two contacts,
including the time and location of ball's bounce (time of ball bounce $(t^i_{b})$
and shot placement position $(\mathbf{x}^i_\mathbf{b})$).
In cases where the second ball contact is a volley,
we adjust the search algorithm to find a trajectory that travels directly between these contact times/locations without a bounce. (The database clip stores the time and location where the ball would have bounced had it not been contacted in the air by the opponent.)
We provide full details of ball trajectory estimation in the supplemental material.

\section{Modeling Player Behavior}
\label{sec:behavior}

\begin{figure}[t]
	\centering
	\includegraphics[width=\columnwidth]{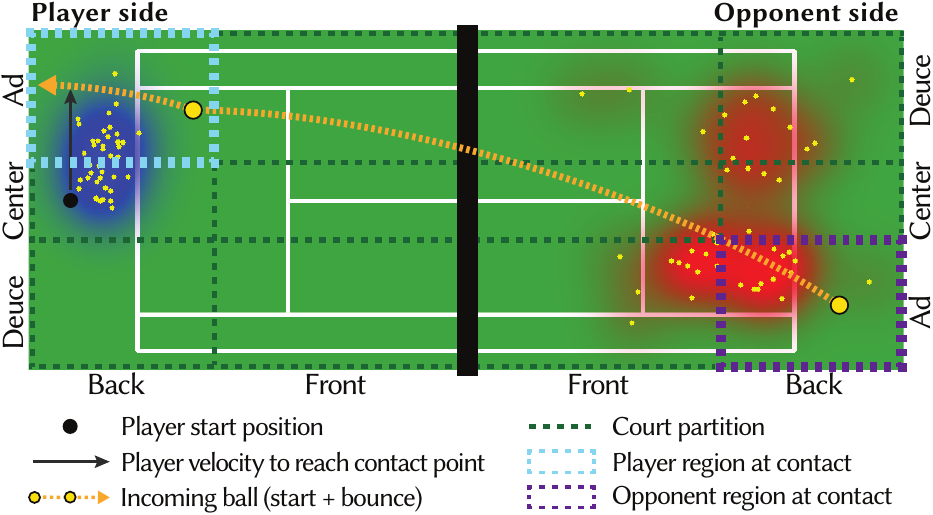}
	\caption{
	Visualization of the shot placement distribution, $p(\mathbf{x_b} | D_s, c=\text{FH-T})$ (red),
	and recovery position distribution, $p(\mathbf{x_r} | D_r, Y=0)$ (blue),
	given a point configuration where the player's estimated ball contact point lies within
	the cyan region,
	the opponent's expected position at ball contact lies within the purple region,
	and the incoming ball follows the given trajectory (orange).
	(FH-T denotes a forehand-topspin shot, $Y=0$ denotes the decision to not approach the net during post-shot recovery.)
	The dotted court boundaries denote the regions used to discretize the positions of players and the incoming ball's starting and bounce positions.
	The velocity required for the player to reach the estimated contact point
	is also discretized to determine $D_s$.
	Yellow dots indicate the shot placement positions and recovery positions
	for database clips that match $D_s$ and $D_r$, 
	and are used to estimate the distributions.}
    \label{fig:behavior_model}
\end{figure}

Using the database of annotated shot cycles,
we construct statistical player behavior models
that input the state of the point at the beginning of the shot cycle,
and produce player shot selection and recovery position decisions for the shot cycle.
We denote this set of behavior decisions as $B^* = (c^*, v^*_b, \mathbf{x^*_b}, \mathbf{x^*_r})$.
(We use $^*$ to denote behavior model outputs.) 
The first three components of $B^*$ refer to shot selection decisions:
shot type $c^*$ (forehand or backhand, topspin or underspin), 
shot velocity $v^*_b$ (the magnitude of average ground velocity from ball contact to bounce) 
and shot placement position $\mathbf{x^*_b}$
(where the shot should bounce on the opponent's side of the court). 
The last component, $\mathbf{x^*_r}$, denotes the player's recovery position goal. 

Tennis point play yields a large space of potential point configurations
that are only sparsely sampled by shot cycle examples in our database.
This prevented use of data hungry models (e.g., neural nets) for modeling player behavior.
To generate realistic behaviors
(follow player specific tendencies, obey tennis principles, model important rare actions)
for a wide range of situations,
we use simple non-parametric models conditioned on a low-dimensional, discretized representation of point state.
We specialize these models and sampling strategies
for the different components of player behavior.

\subsection{Shot Selection}
\label{sec:behavior:placement}

The shot selection behavior model generates $(c^*, v^*_b, \mathbf{x^*_b})$ in $B^*$.
In addition to the positions of both players at the start of the shot cycle 
and the trajectory of the incoming ball,
the shot selection behavior model also utilizes the following
\emph{estimated future state} features that
influence decision making during tennis play.

\emph{Estimated ball contact time/position.}
We estimate the time and court position where the player will contact the incoming ball
by determining when the ball's trajectory intersects the plane (parallel to the net) containing the player.
This computation assumes that a player moves only laterally to reach the ball (not always true in practice),
but this rough estimate of the contact point is helpful for predicting shot placement and recovery behavior.

\emph{Velocity to reach estimated contact position.}
Rapid movement to reach a ball can indicate a player under duress in a rally,
so we compute the velocity needed to move the player from their
position at the start of the shot cycle to the estimated ball contact position. 

\emph{Opponent's estimated position at time of contact.}
A player's shot selection decisions depend on predictions of where their opponent will be in the future.
The opponent's most likely position at the time of player ball contact
is given by the recovery position of the opponent's current shot cycle.  
We use the method described in Section~\ref{sec:behavior:recovery} to estimate this position.

We construct a point state descriptor $D_s$
by discretizing the following five features:
the estimated positions of both players at time of contact, 
the incoming ball's starting and bounce position, 
and the magnitude of player velocity needed to reach the estimated contact point.
We discretize the magnitude of player velocity uniformly into five bins.
Fig.~\ref{fig:behavior_model} visualizes the 2D court partitioning used
to discretize the estimated positions of both players at ball contact
and the incoming ball's starting and bounce positions.
We divide the region between the singles sidelines uniformly into deuce, center, ad regions,
and place the front/back court partitioning line
halfway between the service and baseline
since player movement toward the net past this point typically indicates a transition to hitting volleys. 
The choice of six regions balances the need for sufficient spatial resolution to capture key trends
(like cross court and down-the-line shots, baseline vs. net play)
with the benefits of a low-dimensional state representation for
fitting a behavior model to sparse database examples.
Overall, this discretization yields 1080 unique point states.
(See supplemental for details on the discretization of all features.)  

\paragraph{Model construction}
As an offline preprocess,
we use the clip database to estimate statistical models conditioned on $D_s$.
For each unique state, we identify all database clips that begin from that state,
and use these data points to estimate models for the various shot selection decisions.
(When computing $D_s$ for database clips,
we directly use the known player positions at ball contact instead of estimating these quantities.)
We model shot type as a categorical distribution $p(c | D_s)$
and model shot velocity, $p(v_b | D_s, c)$,
and shot placement, $p(\mathbf{x_b} | D_s, c)$,
using 1D and 2D kernel density estimators (KDE)\,\cite{silverman:2018:density}
conditioned on both $D_s$ and shot type.
(Fig.~\ref{fig:behavior_model} visualizes the $p(\mathbf{x_b} | D_s, c=\text{FH-T})$ in red).
As many point state descriptors correspond to at most a few database clips,
we use a large, constant Gaussian kernel bandwidth for all KDEs.
(We use Scikit-learn's leave-one-out cross-validation to estimate the bandwidth that maximizes the log-likelihood of the source data\,\cite{scikit-learn}.)
Since a player's shot placement depends both on their own playing style 
and also their opponent's characteristics
(e.g. a player may be more likely to hit to their opponent's weaker side),
we build opponent-specific models of player behavior by limiting source data to matches involving the two players in question.

\paragraph{Model evaluation}
To generate a player's shot selection decision during a novel rally,
the system computes $D_s$ from the current state of the point,
then samples from the corresponding distributions to obtain $(c^*, v^*_b, \mathbf{x^*_b})$.
To emphasize prominent behaviors,
we reject samples with probability density lower than one-tenth of the peak probability density.
If the generated shot placement position falls outside the court,
the player's shot will be an error and the point will end (Section~\ref{sec:point_ending}).
Therefore, the shot selection model implicitly encodes the probability
that a player will make an error
given the current point conditions.

The approach described above estimates behavior probability distributions
from database clips that exactly match $D_s$. 
When there is insufficient data to estimate these distributions,
we relax the conditions of similarity until sufficient database examples 
to estimate a distribution are found.  
Specifically, we estimate the marginal distribution
over an increasingly large set of point states,
prioritizing sets of states that differ only in features (variables in $D_s$)
that are less likely to be important to influencing behavior.
We denote $D_s^k$ as the set of point states that differ from $D_s$ by exactly $k$ features.
Our implementation searches over increasing $k$ ($1 \le k \le4$) until a $D_s^k$ is found for which there are sufficient matching database clips (at least one clip in our implementation).
For each $k$, we attempt to marginalize over features in the following order determined from domain knowledge of tennis play: velocity to reach the ball (least important), incoming ball bounce position, incoming ball starting position and lastly the opponent's position at time of contact (most important).  
In practice, we find 91\% of the shot selection decisions made in novel points do not require marginalization.

\subsection{Player Recovery Position}
\label{sec:behavior:recovery}

A player's recovery position reflects expectations about their opponent's next shot.
Since a player's shot selection in the current shot cycle influences their opponent's next shot,
the input to the player recovery position behavior model
is a new descriptor $D_r$ formed by concatenating $D_s$ (the current state of the point)
with the shot placement position (from the shot selection behavior model).

One major choice in recovery positioning is whether the player
will aggressively approach the net following a shot.
We model this binary decision $Y$ as a random variable 
drawn from a Bernoulli distribution $p(Y | D_r)$,
where probability of $Y$=$1$ is given by the fraction of database clips
matching $D_r$ that result in the player approaching the net (recovering to front regions of the court). 
We model $p(\mathbf{x_r} | D_r, Y)$ by constructing a 2D KDE from the recovery position of database clips matching both $D_r$ and $Y$.
(Fig.~\ref{fig:behavior_model} visualizes $p(\mathbf{x_r} | D_r, Y = 0)$ in blue.)
When insufficient matching database clips exist,
we marginalize the distribution using the same approach as for the shot selection behavior model. 

In contrast to shot selection,
where we sample probability distributions to obtain player behavior decisions
(players can make different shot selection decisions in similar situations),
the player behavior model designates the player's recovery position goal as the \emph{most likely}
position given by $p(\mathbf{x_r} | D_r, Y)$.
This reflects the fact that after the strategic decision of whether or not to approach the net is made,
a player aims to recover to the position at the baseline (or the net)
that is most likely to prepare them for the next shot.


\section{Clip Search}
\label{sec:motion}

At the beginning of each shot cycle,
the system must find a database video clip that best 
meets player behavior goals, while also yielding high visual quality.
Specifically,
given the desired player behavior $B^* = (c^*, v^*_b, \mathbf{x^*_b}, \mathbf{x^*_r})$,
as well as the incoming ball's trajectory ($\mathbf{x_{ball}}(t)$),
player's court position ($\mathbf{x_{p0}}$),
pose ($\mathbf{p_0}$), and root velocity ($\mathbf{v_{p0}})$
(computed by differencing the court positions in neighboring frames)
at the start of the shot cycle,
we evaluate the suitability of each database clip using a cost model that aims to:

\begin{itemize}
    \item Make the player swing at a time and court position that places the racket
    in the incoming path of the ball.
    \item Produce a shot that matches type $c^*$, travels with velocity $v^*_b$ and bounces near $\mathbf{x^*_b}$.  
    \item Move the player to $\mathbf{x^*_r}$ after ball contact.
    \item Exhibit good visual continuity with the player's
    current pose $\mathbf{p_0}$ and velocity $\mathbf{v_{p0}}$.
\end{itemize}

Since it is unlikely that any database clip will precisely meet all conditions,
we first describe clip manipulations that improve a clip's match with specified constraints
(Sec.~\ref{sec:motion_reaction}, Sec.~\ref{sec:motion_recovery}), 
then describe the proposed cost model (Sec.~\ref{sec:clip_cost}).
We also cover details of how clip search is used to determine when points end (Sec.~\ref{sec:point_ending}).

\begin{figure}[t]
	\centering
	\includegraphics[width=\columnwidth]{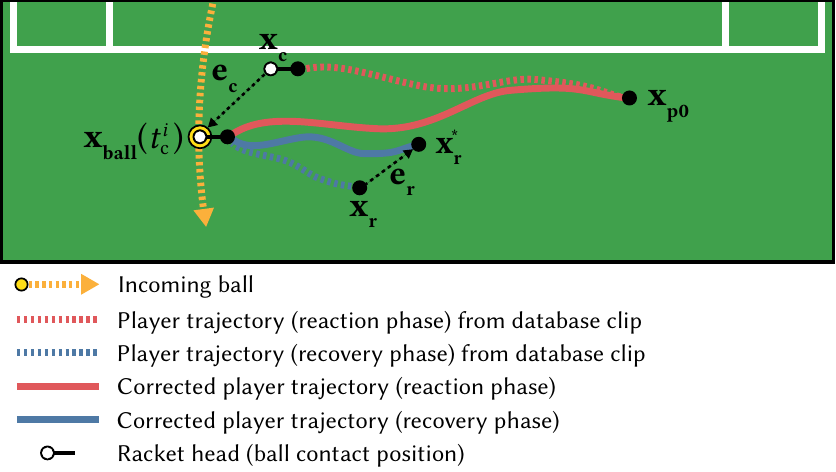}
	\caption{Translational correction in clip search.
		Using motion directly from a database clip would place the player's racket at $\mathbf{x_c}$ at ball contact time, and the player at $\mathbf{x_r}$ at the end of the shot cycle.
		We compute the error in these positions $\mathbf{e_c}$ (the distance from $\mathbf{x_c}$ to the ball)
		and $\mathbf{e_r}$ (distance to the target recovery position $\mathbf{x^*_r}$),
		and apply a translational correction to the root position of the player so that the racket meets the incoming ball,
		and that player reaches the target recovery position at the end of the shot cycle.}
	\label{fig:clip_search}
\end{figure}

\subsection{Making Ball Contact}
\label{sec:motion_reaction}

During the reaction phase of the shot cycle,
the goal is to produce an animation that moves the player from $\mathbf{x_{p0}}$
to a position where their swing makes contact with the incoming ball. 
Rather than specify a specific ball contact point on the incoming trajectory,
we allow the player to hit the ball \emph{at any feasible point on its trajectory}.
For each clip $i$ in the database,
we use the length of the reaction phase to determine the incoming ball's position
at the clip's ball contact time $\mathbf{x_{ball}}(t^i_c)$. 
After translating the player in the clip in court space to start the shot cycle at $\mathbf{x_{p0}}$,
the player's shot contact position (the position of the racket head)
resulting from the clip is given by:
\begin{align}
	\mathbf{x_c} = \mathbf{x_{p0}} + \big(\mathbf{x}^i_{\mathbf{c}}  - \mathbf{x}^i_{\mathbf{p}}(0) \big)
\end{align}
For the rest of the paper, we use superscript $i$ notation to refer to positions/times
from source database clips (Table~\ref{tab:clipschema}).
Variables without a superscript denote positions/times in the synthesized point.

\paragraph{Translational Correction}
Directly using the player's motion from the database clip is unlikely to make the player's swing contact the incoming ball. 
To make the player appear to hit the ball at the clip's ball contact time ($t_c^i$),
we add a court-space translational correction to move the player
so that the racket head is located at the same location as the ball at the time of contact 
(Fig.~\ref{fig:clip_search}).
We compute the position error in the racket head:
\begin{equation}
\mathbf{e_c} = \mathbf{x_{ball}}(t^i_c) - \mathbf{x_c}
\label{eqn:contact_error}
\end{equation}
and modify the player's motion in the clip with a court space translational correction that places the player at the correct spot on the court to hit the ball.
This correction is increasingly applied throughout the reaction phase by linearly
interpolating the uncorrected and corrected motions.
This yields a corrected player court space position for the reaction phase:
\begin{align}
\mathbf{x_p}(t) = \mathbf{x_{p0}} + \big( \mathbf{x}^i_{\mathbf{p}}(t)  - \mathbf{x}^i_{\mathbf{p}}(0) \big) + w\left(\frac{t}{t^i_c} \right) \mathbf{e_{c2d}}
\label{eqn:player_position}
\end{align} 
where $w(x)$ is an ease-in, ease-out function used to generate interpolation weights, and
$\mathbf{e_{c2d}}=[ \mathbf{e_{c}}.x,\quad \mathbf{e_{c}}.y, \quad 0 ]^T$.
Since translational correction does not move the player off the ground plane,
it does not eliminate the $z$ component (height) of contact point error. 

A large translational correction can result in objectionable foot skate,
so clip search aims to minimize $\mathbf{e_{c2d}}$ (Sec.~\ref{sec:clip_cost}).
If there is no clip in the database with a sufficiently small translational correction,
the player cannot reach the ball, signaling the end of the point (Sec.~\ref{sec:point_ending}).

\subsection{Meeting Recovery Constraints}
\label{sec:motion_recovery}

For the recovery phase of the shot cycle,
we compute a translational correction to move the player's position at the end of the recovery phase $\mathbf{x_p}(t_r^i)$ to the target recovery position $\mathbf{x^*_r}$.
This correction is computed similarly to the correction used to ensure ball contact (Fig.~\ref{fig:clip_search}): 

\vspace{-.5em}
\begin{align}
\mathbf{x_r} &= \mathbf{x_p}(t^i_c) + \big( \mathbf{x}^i_{\mathbf{p}}(t^i_r) - \mathbf{x}^i_{\mathbf{p}}(t^i_c) \big) 
\label{eqn:player_position} \\ 
\mathbf{e_r}     &= \mathbf{x^*_r} - \mathbf{x_r}
\end{align}

Note from the pseudocode in Fig.~\ref{fig:overview} that the actual length of the recovery phase of 
the shot cycle initiated in this iteration is not known when clip search occurs (line 9).  
The length 
is not determined until the next loop iteration,
when clip search determines when the opponent will hit the return shot.
Therefore, when evaluating clip cost during search,
we assume the length of the new shot cycle's recovery phase is the same as that of the database clip ($t^i_{r}-t^i_{c}$) when computing $\mathbf{e_r}$.
When the opponent's future ball contact time is determined (in the next loop iteration),
we compute the actual $\mathbf{e_r}$ for the chosen clip
by replacing $t^i_r$ with this time in Eqn.~\ref{eqn:player_position} (Fig.~\ref{fig:overview}, line 10).
The actual $\mathbf{e_r}$ is used to compute the translational correction
used for recovery phase sprite rendering.

To avoid excessive foot skate during recovery,
we constrain $\mathbf{e_r}$ to a visually acceptable range
(see $\mathbf{C_{recover}}$, Sec.~\ref{sec:clip_cost}).
Therefore, unlike the reaction phase, where translation is used to bring the player's racket exactly into the trajectory of the incoming ball to continue the point, the recovery translation may not always move the player exactly to the target recovery position $\mathbf{x^*_r}$.
We judged that failing to precisely meet the recovery goal
was acceptable in times when doing so would create notable visual artifacts.   

\subsection{Clip Cost}
\label{sec:clip_cost}
We compute the cost of each database clip as a weighted combination of terms that assess the continuity of motion between clips,
the amount of manipulation needed to ensure ball contact and meet recovery position goals,
as well as how well a clip matches the shot velocity and placement components of shot selection goals.

\vspace{0.5em}
\noindent
\textbf{Pose match continuity.} ($\mathbf{C_{pose}}$) We compute the average screen-space $L2$ distance of joint positions between the current pose (the last frame of the previous shot cycle) $\mathbf{p_0}$ and the starting pose of the database clip $\mathbf{p}^i(0)$.
Since the size of a player's screen projection depends on their court location,
we scale $\mathbf{p}^i(0)$ by a factor $\sigma$ computed from the court positions of the two poses to normalize the screen-space size of poses (details about $\sigma$ in Sec.~\ref{sec:rendering_runtime}). 
\begin{align*}
	\mathbf{C_{pose}} = \| \mathbf{p_0} - \sigma \mathbf{p}^i(0) \|
\end{align*}

\vspace{0.5em}
\noindent
\textbf{Player root velocity continuity.} ($\mathbf{C_{velo}}$). We compute the difference between the player's current velocity $\mathbf{v_{p0}}$ and the initial velocity of the database clip $\mathbf{v}^i_\mathbf{p}(0)$. ($\mathbf{v}^i_\mathbf{p}(t)$ is computed by differencing $\mathbf{x}^i_\mathbf{p}(t)$ in neighboring frames.)
\begin{align*}
	\mathbf{C_{velo}} = \| \mathbf{v_{p0}} - \mathbf{v}^i_\mathbf{p}(0) \|
\end{align*}

\vspace{0.5em}
\noindent
\textbf{Ball contact height matching.} ($\mathbf{C_{contact}}$) We use this term to force the player's racket height at contact to be close to the ball's height. (Recall that the correction $\mathbf{e_{c2d}}$ only corrects error in the racket's XY distance from the ball.)
\begin{align*}
	\mathbf{C_{contact}} = \| \mathbf{x_c}.z - \mathbf{x_{ball}}(t_c^i).z \|
\end{align*}

\vspace{0.5em}
\noindent
\textbf{Reaction phase correction.} ($\mathbf{C_{react}}$).
This term assigns cost to the translational correction applied to make the player contact the ball.  
We observe that high velocity clips can undergo greater correction without noticeable artifacts,
so we seek to keep velocity change under a specified ratio.
Also, translations that significantly change the overall direction of a player's motion in the source clip are easily noticeable,
so we penalize translations that create these deviations.
This results in the following two cost terms:

\begin{align*}
& \mathbf{C_{react, velo}} = 
\max(
\frac{\| \Delta\mathbf{x_{cor}} \|}
	 {\| \Delta\mathbf{x_{db}} \|},
\frac{\| \Delta\mathbf{x_{db}} \|}
	 {\| \Delta\mathbf{x_{cor}} \|}
)\\
& \mathbf{C_{react, dir}} = 
1 - \cos(\Delta\mathbf{x_{db}}, \Delta\mathbf{x_{cor}})
\end{align*}

Where
$\Delta\mathbf{x_{db}} = \mathbf{x}_{\mathbf{p}}^i(t_c^i) -
\mathbf{x}_{\mathbf{p}}^i(0)$ and 
$\mathbf{\Delta x_{cor}} = \mathbf{x}_{\mathbf{p}}(t_c^i) - \mathbf{x_{p0}}$
give the player's uncorrected and translation corrected movement during the reaction phase of the database clip. 

\vspace{0.5em}
\noindent
\textbf{Recovery phase correction.} ($\mathbf{C_{recover}}$).  The term measures the amount of correction needed to achieve the target player recovery position, and it is computed in the same way as $\mathbf{C_{react}}$.

\vspace{0.5em}
\noindent
\textbf{Shot match.} ($\mathbf{C_{shot}}$)  This term ensures that the chosen clip matches the target shot type $c^*$ and also measures how closely the shot velocity $v_b^i$ (derived from the estimated ball trajectory) and placement position $\mathbf{x}^i_\mathbf{b}$ of the shot in the database clip match the behavior goals $v^*_b$ and $\mathbf{x^*_b}$ (after accounting for player translation).  
\begin{align*}
& \mathbf{C_{shot, type}} = \begin{cases}
	0 & \text{if $c^i =c^*$} \\
	\inf & \text{otherwise}
	\end{cases} \\
& \mathbf{C_{shot, velo}} = \| v_b^i - v^*_b\| \\
& \mathbf{C_{shot, place}} = \| \mathbf{x^*_b} - \big(\mathbf{x}^i_\mathbf{b} + \big( \mathbf{x_p}(t^i_c) - \mathbf{x}^i_{\mathbf{p}}(t^i_c) \big) \big) \|
\end{align*}

The total cost is computed as a weighed combination of the aforementioned cost terms.
We give the highest weight to $\mathbf{C_{recover, dir}}$ since recovering in the wrong direction yields extremely implausible player behavior.
We also prioritize the weights for $\mathbf{C_{react, velo}}$ and $\mathbf{C_{react, dir}}$
to reduce foot skate during the reaction phase.
We give lower weight to $\mathbf{C_{contact}}$ since the fast motion of a swing (the racket is motion blurred) makes it challenging to notice mismatches between the position of the ball and racket head at contact,
and to  $\mathbf{C_{shot, place}}$ since inconsistencies between swing motion and
the direction the simulated ball travels typically require tennis expertise to perceive.
(See supplemental for full implementation details.)

\subsection{Point Ending}
\label{sec:point_ending}

In our system a point ends when one of two events occur:
a player makes an error by hitting the ball outside the court,
or a player hits a shot that lands in the court but cannot be reached by the other player.  
The shot selection behavior model determines when errors occur (Section~\ref{sec:behavior:placement}).
We use clip search and the amount of required recovery phase translation correction
to determine when a player cannot reach an incoming ball.  
If no clip in the database has sufficiently low $\mathbf{C_{react}}$,
then the player cannot reach the ball and the system determines they lose the point.

The point synthesis algorithm
determines that the simulated point will end in the current shot cycle \emph{before}
performing clip search (Fig.~\ref{fig:overview}, line 8).
This allows database clips to
be excluded from clip search if they do not match the desired shot cycle outcome
(as determined by the clip's $o^i$).
For example, when the current shot does not cause the point to end,
clip search is limited to database clips that are not point ending shot cycles.
These clips depict the player rapidly returning to a ready position,
poised to react to the next shot.
If the player cannot reach the ball,
search is limited to only database clips with the same outcome,
resulting in output where the player concedes the point without making ball contact.
Finally, since there are only a small number of clips in the database depicting point ending shots
(at most one clip per point), 
if the player will make an error or hit a winner,
we search over all clips where ball contact is made (both point ending and not point ending),
choosing one that minimizes the total cost.
Although not shown in Fig.~\ref{fig:overview}'s pseudocode,
if the best clip does not depict a point ending shot cycle,
after ball contact we insert an additional clip transition
to the recovery phase of a point ending clip
to depict the player stopping play at the end of the final shot cycle.


\section{Rendering}
 \label{sec:rendering}

Given the court space position ($\mathbf{x_p}(t)$ from Eqn.~\ref{eqn:player_position}),
of a player and the video clip selected to depict the player during a shot cycle,
we render the player by compositing sprites (players with rackets) from the source clip
onto the court background image (a frame chosen from one of the broadcast video) with generated alpha mattes.
Achieving high output quality required overcoming several challenges that arise when using broadcast video as input: 
maintaining smooth animation in the presence of imperfect (noisy) machine annotations,
hallucinating missing pixels when player sprites are partially cropped in the frame,
and eliminating variation in player appearance across source clips.

\subsection{Player Sprite Rendering}
\label{sec:rendering_runtime}

\begin{figure}[t]
	\centering
	\includegraphics[width=\columnwidth]{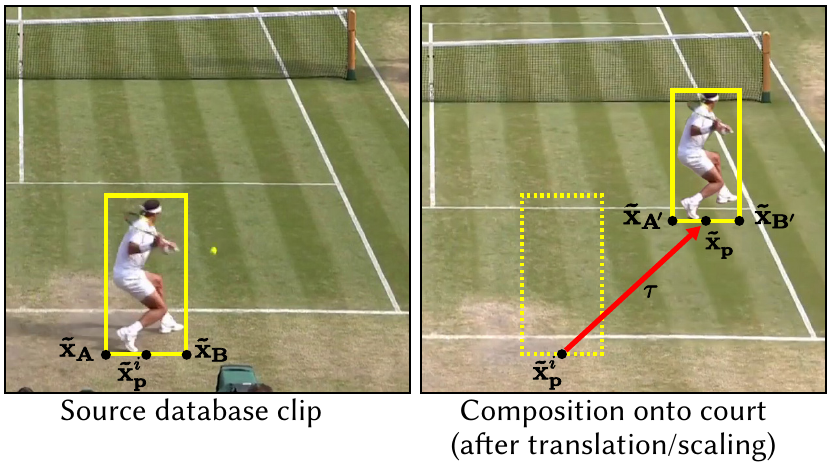}
	\caption{Source video sprites are translated and scaled to render the player at a novel location in court space.  Scaling requires an estimate of the court space extent of the player's bounding box, which is computed from projection of the boxes' bottom-left and bottom-right screen space points ($\mathbf{\tilde{x}_{A}},\mathbf{\tilde{x}_{B}}$). }
	\label{fig:trans_scaling}
\end{figure}

\paragraph{Sprite Transformation.} 
We translate and scale the source clip sprite to emulate perspective projection onto the court (Fig.~\ref{fig:trans_scaling}).
Given a source clip $i$ with player bounding box (bottom center) located at root position $\mathbf{\tilde{x}}^i_\mathbf{p}(t)$ (notations with $\tilde{}$ refer to positions in screen space), 
we compute the screen space translation, $\mathbf{\tau}(t)$, of source pixels to the target court background image as:
\begin{align*}
	\mathbf{\tau}(t) = \mathbf{H}^{-1}\mathbf{x_p}(t) - \mathbf{\tilde{x}}^i_\mathbf{p}(t) 
\end{align*}
where $\mathbf{H}^{-1}$ is the inverse homography mapping between the court space and screen space in the background image. 

To determine the sprite's scaling due to perspective projection,
we use the bottom-left $\mathbf{\tilde{x}_{A}}$ and bottom-right $\mathbf{\tilde{x}_{B}}$ points
of the source clip bounding box to compute its court space extent (using $\mathbf{H}^i(t)$ for the source frame),
translate this segment (in court space) so that it is centered about
the player's current position $\mathbf{x_p}(t)$,
then compute the screen space extent of the translated segment
$||\mathbf{\tilde{x}_{B'}}-\mathbf{\tilde{x}_{A'}}||$.
The ratio of the horizontal extent of the source clip segment and translated segment
determines the required scaling, $\sigma(t)$, of the sprite.

\begin{align*}
	\mathbf{\tilde{x}_{A'}} &= \mathbf{H}^{-1} \Big( \mathbf{H}^i(t)\mathbf{\tilde{x}_{A}} +\big(\mathbf{x_p}(t) - \mathbf{x}^i_{\mathbf{p}}(t) \big) \Big) \\
	\mathbf{\tilde{x}_{B'}} &= \mathbf{H}^{-1} \Big( \mathbf{H}^i(t)\mathbf{\tilde{x}_{B}} + \big(\mathbf{x_p}(t) - \mathbf{x}^i_{\mathbf{p}}(t) \big) \Big)\\	
	\sigma(t) &= \frac{||\mathbf{\tilde{x}_{B'}}-\mathbf{\tilde{x}_{A'}}||}{||\mathbf{\tilde{x}_{B}}-\mathbf{\tilde{x}_{A}}||}
\end{align*}

Due to noise in estimates of a player's 
root position $\mathbf{\tilde{x}}^i_\mathbf{p}(t)$,
performing the above translation and scaling calculations each frame
yields objectionable frame-to-frame jitter in rendered output.
To improve smoothness, 
we compute $\mathbf{\tau}(t)$ and $\sigma(t)$ only at 
the start of the reaction phase,
ball contact time, and the end of recovery phase,
and interpolate these values to estimate sprite translation
and scaling for other frames in the shot cycle.

\paragraph{Clip Transitions}
To reduce artifacts due to discontinuities in pose at clip
transitions, we follow prior work on video textures\,\cite{Flagg:2009:humanvideotex}
and warp the first five frames of the new shot cycle clip
towards the last frame of the prior clip using moving least
squares\,\cite{Schaefer:2006:movingleast}.
Linear interpolation of pose keypoints from the source and destination
frames are used as control points to guide the image warping.
We produce final frames by blending the warped frames of the
new clip with the last frame of the prior clip.

\paragraph{Matte Generation}
To generate an alpha matte for compositing sprites into the background image of the tennis court, 
we use Mask R-CNN\,\cite{He:2017:maskrcnn} to produce a binary segmentation of
pixels containing the player and the racket in the source clip frame,
then erode and dilate\,\cite{opencv_library} the binary segmentation mask to create a trimap.
We then apply Deep Image Matting\,\cite{Xu:2017:deepmatting} to the original frame and the trimap
to generate a final alpha matte for the player.
Matte generation is performed offline as an preprocessing step, so it incurs no runtime cost.
In rare cases, we manually correct the most eggregious segmentation errors made by this automatic process.
(We corrected only 131 frames appearing in the supplemental video.)

\paragraph{Player Shadows}
We generate player shadows using a purely image-based approach that warps the player's alpha matte to approximate projection on the ground plane from a user-specified lighting direction. The renderer darkens pixels of the background image that lie within the warped matte. (See supplemental for a description of the control points used for this image warp.) This approach yields plausible shadows for the lighting conditions in our chosen background images, and avoids the challenge of robust 3D player geometry estimation.

\begin{figure}[t]
	\centering
	\includegraphics[width=\columnwidth]{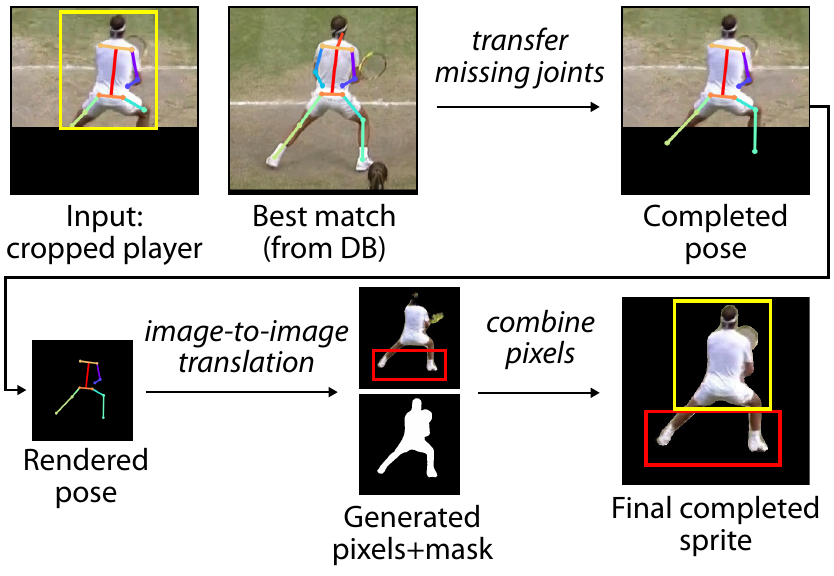}
	\caption{When the player in an input clip lies partially outside the frame,
		we hallucinate missing pose keypoints from similar (fully visible) sprites,
		then use paired image translation to hallucinate missing pixels from the completed skeleton.
		In this example, the final sprite contains a combination of original pixels and hallucinated legs.}
	\label{fig:croppingfix}
\end{figure}

\begin{figure*}[t]
	\centering
	\includegraphics[width=\textwidth]{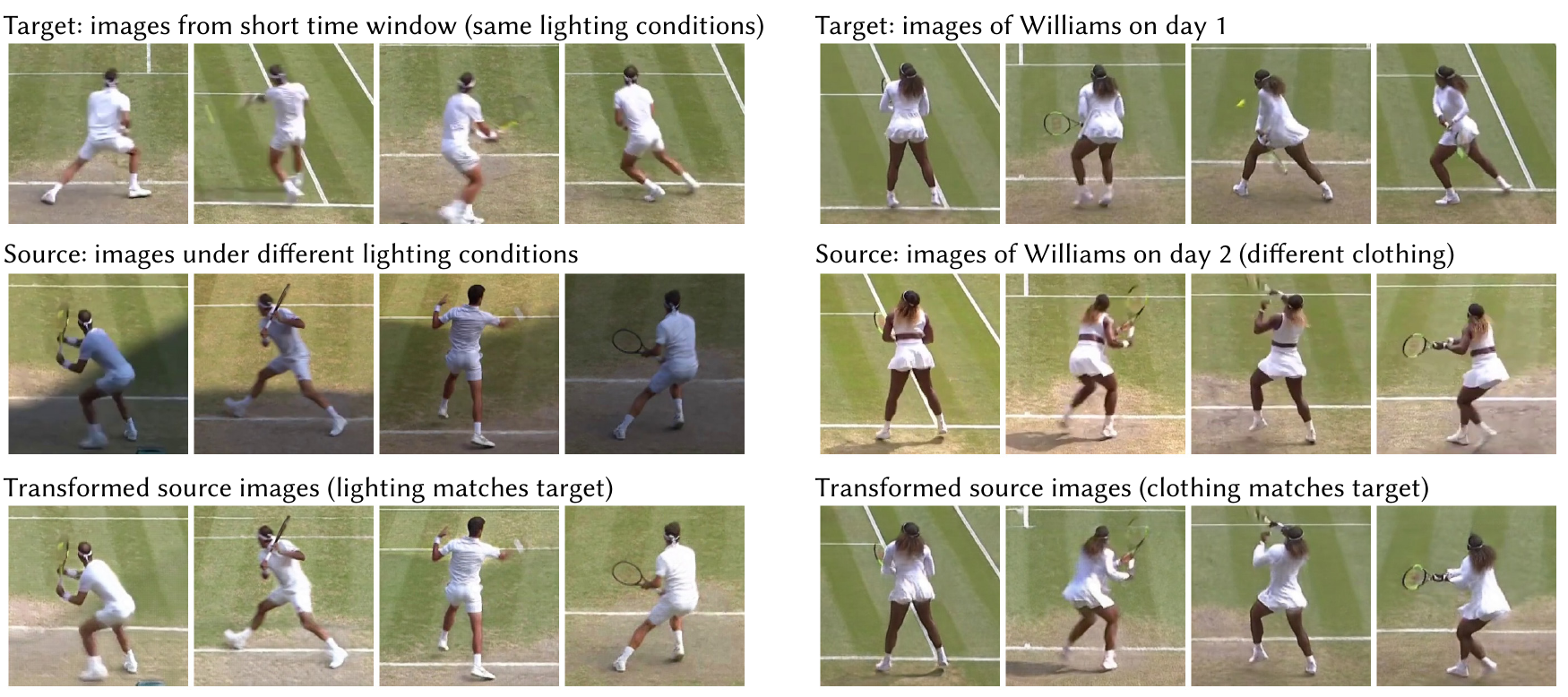}
	\caption{
		We train CycleGAN models\,\cite{Zhu:2017:cycleGAN} to perform unpaired image-to-image translation to reduce differences in a player's appearance across video clips.
		These corrections enable a larger number of video clips to be used as inputs for video sprite synthesis.
		Left:  removing lighting differences.
		Right: modifying Serena Williams' clothing (adding long sleeves) to match her outfit on a prior day.}
	\label{fig:normalization}
\end{figure*}

\subsection{Appearance Normalization and Body Completion}
\label{sec:rendering_preprocess}

Using unmodified pixels from database player sprites results in
unacceptable output quality due to appearance differences across clips tand missing pixel data. 
To reduce these artifacts, we perform two additional preprocessing steps that manipulate the appearance
of source video clips to make them suitable for use in our clip-based renderer.
Both preprocessing steps modify the pixel contents of video clips used by the renderer.
They incur no additional run-time performance cost.

\paragraph{Hallucinating Cropped Regions}
When the near court player in the source video is positioned far behind the baseline,
part of their lower body may fall outside the video frame (Fig.~\ref{fig:croppingfix}, top-left).
In fact, 41\%, 39\%, 15\%, 9\% of the near court clips for Nadal, Djokovic, Federer and Serena Williams suffer from player cropping. 
To use these cropped clips in other court locations we must hallucinate the missing player body pixels, otherwise rendering will exhibit visual artifacts like missing player feet.
We identify clips where player cropping occurs using the keypoint confidence scores from pose detection for the near court player
and classify clips by the amount of body cropping that occurs:
no cropping, missing feet,
and substantial cropping (large parts of the legs or more are missing).
We discard all clips with substantial cropping,
and hallucinate missing pixels for clips with missing feet using the process below.

For each frame containing cropped feet,
we find an uncropped frame of the same player in the database that features the most similar pose (Fig.~\ref{fig:croppingfix}, top-center).
We use the average L2 distance of visible pose keypoints as the pose distance metric.
We transfer ankle keypoints from the uncropped player's pose to that of the cropped player.
(In practice we find the similar pose match in $N$=5 consecutive frames from the uncropped clip for temporal stability.)
Once the positions of the missing keypoints are known,
we follow the method of\,\cite{Chan:2019:everybodydance} by rendering a stick skeleton figure
and using paired image-to-image translation to hallucinate a player-specific image of the missing body parts (Fig.~\ref{fig:croppingfix}, bottom).
The matte for the hallucinated part of the player is determined by the background (black pixels) in the hallucinated output.
We construct the final uncropped sprite by compositing the pixels from the original image
with the hallucinated pixels of legs and feet.

\paragraph{Normalizing Player Appearance}
Our database contains video clips drawn from different days and different times of day,
so a player's appearance can exhibit significant differences across clips. 
For example,
the direction of sunlight changes over time (Fig.~\ref{fig:normalization}-left),
the player can fall under shadows cast by the stadium, and
a player may wear different clothing on different days.
(Serena Williams wears both a short sleeved and a long sleeved top, Fig.~\ref{fig:normalization}-right.)
As a result, transitions between clips can yield jarring appearance discontinuities.

Eliminating appearance differences across clips using \emph{paired} image-to-image translation (with correspondence established through a pose-based intermediate representation\,\cite{Chan:2019:everybodydance,kim:2018:deepvideoportrait}) was unsuccessful
because pose-centric intermediate representations fail to capture dynamic aspects of a player's appearance,
such as wrinkles in clothing or flowing hair.
(Paired image-to-image translation was sufficient to transfer appearance for the players feet, as described in the section above, because appearance is highly correlated to skeleton in these body regions.)

To improve visual quality,
we use \emph{unpaired} image-to-image translation for appearance normalization (Fig.~\ref{fig:normalization}).
We designate player crops from a window of time in one tennis match as a target distribution ($320\times320$ pixel crop surrounding the center point of the player's bounding box),
then train CycleGAN\,\cite{Zhu:2017:cycleGAN} to translate other player crops into this domain.
To correct for lighting changes, we train a CycleGAN for each of our source videos to translate player
crops from different times of the day to a distribution obtained from the first hour of the 2019 Federer-Nadal Wimbledon semi-final, from which we also pick a frame without players as our court background to ensure lighting consistency. 
When distribution shift in player appearance is due to changes of clothing across matches,
we train CycleGAN models to transfer player appearance between specific matches.


\section{Results and Evaluation}
\label{sec:eval}

\begin{table}[t]
	\resizebox{\linewidth}{!}{
		\centering
		\begin{tabular}{l l r r r r r}
			Task & Model  & Total (hr)   & Per frame (ms)\\
			\hline
			Bbox + keypoint   & Detect-and-track\,\cite{Girdhar:2018:detecttrack}	& 25   & 480 \\
			\hline
			Segmentation      & Mask R-CNN\,\cite{He:2017:maskrcnn} 		 	    & 572  & 11200 \\
			\hline
			Matting      	  & Deep Image Matting\,\cite{Xu:2017:deepmatting} 		& 36   & 350 \\
			\hline
			Body Completion   & Pix2PixHD\,\cite{Wang:2018:pix2pixHD} 		 		& .05  & 10 \\
			\hline
			Appearance Norm   & CycleGAN\,\cite{Zhu:2017:cycleGAN}  		 		& 35   & 270 \\
			\hline
		\end{tabular}
	}
	
	\caption{Offline preprocessing times (both per frame, and for all frames (total), on a TITAN~V GPU) 
		for operations needed to annotate database video clips and prepare them for rendering.}
	\label{tab:runtimetable}
\end{table}

\begin{figure*}[t]
	\centering
	\includegraphics[width=\textwidth]{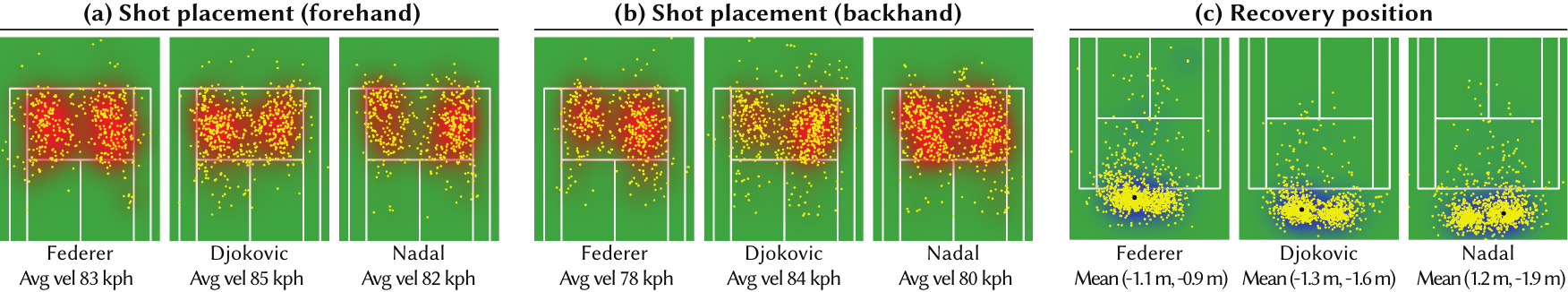}
	\caption{Shot placement (forehand/backhand) and recovery positions of Federer (right-handed), Djokovic (right-handed), and Nadal (left-handed) in simulated points.
		(The red and blue heat maps visualize a KDE fit to the results of simluation,
		not the distribution used by our behavior models.)
		In general a majority of shots are hit cross court,
		except for Nadal's backhand which is distributed equally around the court.
		Federer stands nearly one meter closer to the baseline (only 0.9~m behind baseline) than Djokovic (1.6~m) and Nadal (1.9~m),
		suggesting more aggressive play.
		All players recover to an off-center court position,
		making it harder for their opponents to hit the ball to their backhand side. }
	\label{fig:player_style}
\end{figure*}

We implemented our controllable video sprite generation system on the Wimbledon
video database described in Sec.~\ref{sec:dataset}.
On average, video clip search over our largest player database (Djokovic),
takes 685~ms on a single core CPU.
This implementation (in Python) is unoptimized,
and could run in real-time with additional engineering.

Table~\ref{tab:runtimetable} shows the offline preprocessing
time needed to annotate or prepare clips for rendering. We give results on a single TITAN~V GPU but 
use frame-parallel processing across many GPUs to reduce overall preprocessing latency.
Our implementation uses publicly available models for
detect-and-track\,\cite{Girdhar:2018:detecttrack},
Mask R-CNN\,\cite{He:2017:maskrcnn},
and Deep Image Matting\,\cite{Xu:2017:deepmatting}.
We train Pix2PixHD\,\cite{Wang:2018:pix2pixHD} and CycleGAN\,\cite{Zhu:2017:cycleGAN}
architectures for body completion and appearance normalization. (See the supplemental material for details.)
In addition to these machine annotations, it took approximately 10 human-hours to annotate shot contact frames, shot types, and shot outcomes in all 16 hours of the Wimbledon dataset.

\subsection{Emergent Player Behavior}

Our player behavior models produce points that echo real-world player tendencies and advanced tennis play.
To evaluate emergent behavior,
we generate novel points using the three male tennis players in our database
(200 points for each of the three pairings: Djokovic vs. Federer, Djokovic vs. Nadal, Federer vs. Nadal). 
We summarize our observations from these simulated points here,
but refer the reader to the supplementary video for a detailed inspection of results.
(Note that in this section,
mentions of a specific player's behavior refer to the actions of video sprites in simulated points,
not the behavior of the real-life player as observed in database clips.)

\paragraph{Recreation of player style.}
Our behavioral models capture elements of player-specific style that are consistent with
well-known tendencies of the real-life players.
For example, on average Federer stands one meter closer to the baseline than Djokovic and Nadal,
suggestive of his attacking style (Fig.~\ref{fig:player_style}-c).
Federer also hits forehands with significantly higher ground velocity than backhands,
while his opponents exhibit greater balance between their forehand and backhand sides.  
Notice how on average,
all players recover to an off-center position on the court
(right-handed Federer and Djokovic to the left side, and left-handed Nadal to the right).
This positioning makes it more difficult for their opponents to hit shots to their weaker backhand side.
Cross-court shots are the most common shots by all players in the simulated rallies
(Fig.~\ref{fig:player_style}-a/b), echoing statistics of real play (cross court shots have lower difficulty).
Nadal hits a particularly high percentage of his forehands cross court.
Conversely, Nadal's backhand shot placement is the most equally distributed to both sides of the court.

\begin{figure*}[t]
	\centering
	\includegraphics[width=\textwidth]{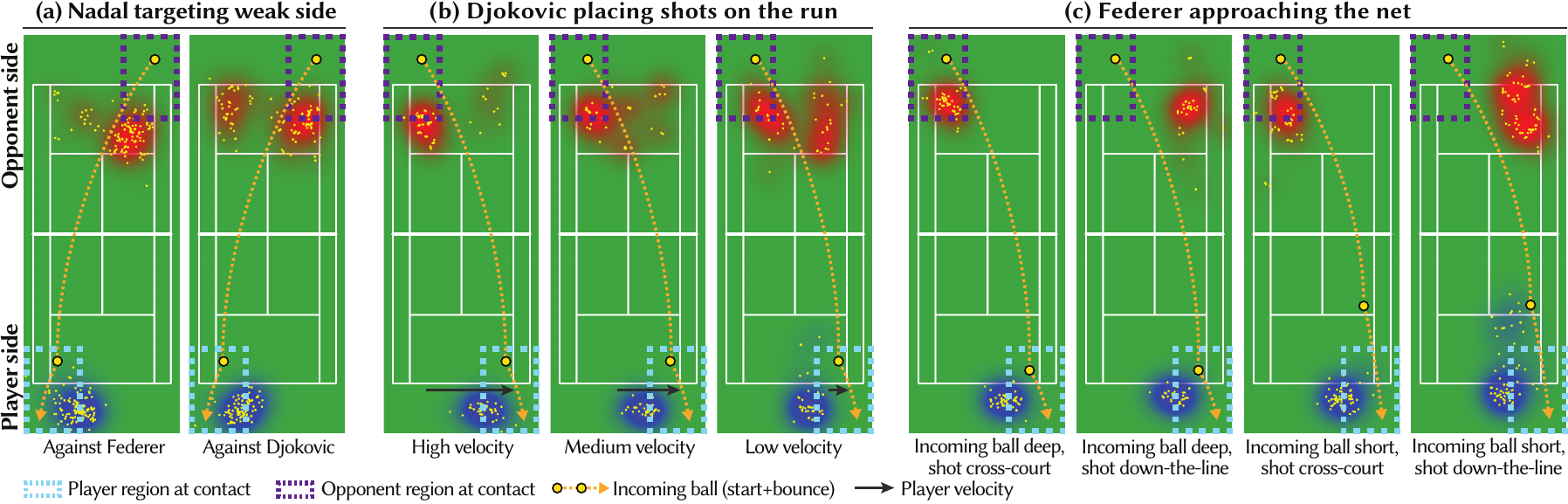}
	\vspace{-5mm}
	\caption{Simulated points feature emergent player behavior that is consistent with good tennis principles.
		(The red and blue heat maps visualize a KDE fit to the results of simluation,
		not the distribution used by our behavior models.)
		(a) When hitting the ball from left side of the court, simulated Nadal places most of his shot cross-court to
		Federer's significantly weaker backhand side.  When playing Djokovic, Nadal directs more shots down the line. 
		(b) When simulated Djokovic has to move rapidly (to the right) to hit an incoming shot vs. Nadal, he plays defensively, placing most of his shots cross court even though that is where Nadal is standing.  When Djokovic does not need to move quickly to reach the incoming ball, he plays more aggressively, placing a greater fraction of shots down the line to the open court.
		(c) Most of the time when simulated Federer approaches the net is after hitting a short incoming ball down the line to the open court.  Positioning oneself at the net following a down the line shot is a common attacking strategy in tennis.}
	\label{fig:emergent_behavior}
\end{figure*}

\paragraph{Playing to opponent's weaknesses.}
In competition, a player's behavior is heavily influenced by the strengths and weakness of their opponent. 
Fig.~\ref{fig:teaser} illustrates differences in Novak Djokovic's shot placement when hitting from the left side of the court when playing Roger Federer (left) and Rafael Nadal (right).
Djokovic places most of his shots cross court to Federer's weaker backhand side.
In contrast, when playing left-handed Nadal,
Djokovic places a majority of his shots down the line to Nadal's backhand.  
Videos generated using these models reflect these decisions.
For example, Djokovic-Federer videos contain many cross-court rallies.
Fig.~\ref{fig:emergent_behavior}-a shows how Nadal plays Federer and Djokovic differently
even though both opponents are right handed.
Nadal avoids Federer's exceptionally strong forehand side
(left side of the figure), but is more willing to hit the ball to Djokovic's forehand.

\paragraph{Players demonstrate good tennis principles.}
Compared to random play, the simulated players play the game according to good tennis principles.
As is the case in real play,
the shot placement of simulated players is affected by whether they are under duress in a rally.
For example, Djokovic's shot placement changes against Nadal,
depending on how fast Djokovic must run to the right to hit the incoming cross-court ball
(Fig.~\ref{fig:emergent_behavior}-b).
When Djovokic must run quickly because he is out of position (left panel),
he typically hits a safer (defensive) cross-court shot.
However, when Djokovic is well positioned to hit the incoming ball (right panel),
he has time to aggressively place more shots down the line to the open court.
We call attention to rallies in the supplemental video where players ``work the point'',
moving the opponent around the court.

In tennis, approaching the net is an important strategy,
but requires the player to create favorable point conditions for the strategy to be successful.
Fig.~\ref{fig:emergent_behavior}-c
visualizes Federer's recovery position after hitting incoming cross-court shots from Djokovic.
Notice that most of the time when Federer approaches the net is 
in situations where the incoming shot bounces in the front regions of the court (incoming shots that land in these regions are easier to hit aggressively)
and when Federer places his shot down the line (allowing him to strategically position himself to cover his opponent's return shot).
Approaching the net after hitting a shot down the line (but not after a cross-court shot) is well known tennis strategy.

Overall, we believe our system generates tennis videos that are substantially more realistic than prior data-driven methods for tennis player synthesis\,\cite{Lai:2012:TennisRealPlay,gafni:2019:vid2game}.
Most notably, compared to recent work on GAN-based player synthesis\,\cite{gafni:2019:vid2game},
our players do not exhibit the twitching motions exhibited by the GAN's output,
make realistic tennis movements (rather than shuffle from side to side),
and perform basic tennis actions, such as making contact with the incoming ball (they do not swing randomly at air) and recovering back into the court after contact.  
We refer the reader to the supplementary video to compare the quality of our results against these prior methods.

\subsection{User Study}



To evaluate whether our player behavior models capture the tendencies
of the real professional tennis players, we conducted a user study
with five expert tennis players.  All participants were current or former
collegiate-level varsity tennis players with at least 10+ years of
competitive playing experience. None of the study participants were
involved in developing our system.

We asked each participant to watch videos of points between Roger
Federer, Rafael Nadal, and Novak Djokovic, generated by our system and
answer questions (on a 5-point Likert scale) about whether a virtual
player's shot selection and recovery point behaviors were
representative of that player's real-life play against the given
opponent. Each participant viewed 15 videos, five videos for each of
three conditions:

\vspace{0.05in}
\noindent
{\bf \em No behavior model.} This configuration generates the
points without using our behavioral model. Specifically, we remove the
shot match and recovery phase correction terms, ($\mathbf{C_{shot}}$ and
$\mathbf{C_{recover}}$) from the cost computation during
motion clip search.
The only behavior constraint in this case is that clips chosen by our system must make ball contact.

\vspace{0.05in}
\noindent
{\bf \em Our behavior model.} This configuration sets behavior targets using our full behavior model as described in Section~\ref{sec:behavior}.  The behavior model is specialized to
each matchup so that the distributions estimated for player behaviors are based on the data we have for the two players involved in the point.

\vspace{0.05in}
\noindent
{\bf \em Oracle.} This configuration sets behavior targets to 
locations observed in real points from broadcast videos.  
In other words, we use our system to
reconstruct a point from our database, directly using the database clips of the point
with their original shot selection and recovery position, 
rather than querying our behavior model.

\begin{figure}[t]
	\centering
	\includegraphics[width=\columnwidth]{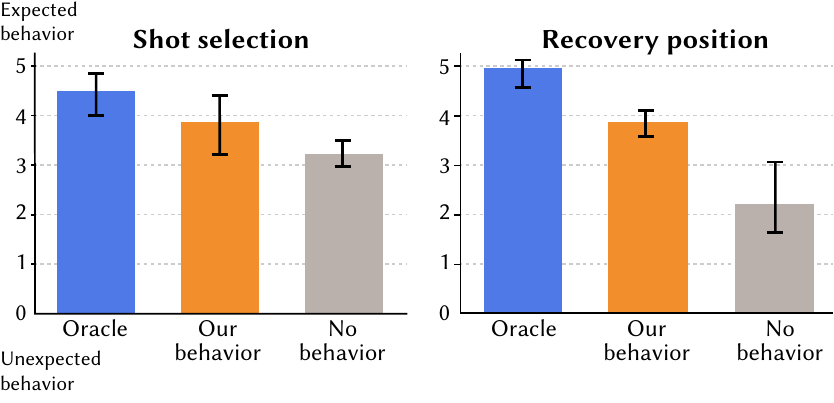}
	\caption{Participants rated our shot selection behavior model ($\mu=3.84, \sigma=0.83$) better than no behavior model ($\mu=3.2, \sigma=0.89$) but less realistic than the oracle ($\mu=4.48, \sigma=0.70$).
		We find all of these differences to be significant running Friedman's non-parametric test ($p<0.001, \chi^2=24.33$) followed by pairwise comparisons using Wilcoxon's signed rank tests ($p<0.019, Z=44.50$ ours vs. no behavior, $p<0.005, Z=24.00$ ours vs. oracle).
		Similarly participants rated our recovery behavior model ($\mu=3.76, \sigma=0.76$) better than no behavior model ($\mu=2.16, \sigma=1.12$) but less realistic than the oracle ($\mu=4.84, \sigma=0.37$).
		Again we find all these differences to be significant running Friedman's test ($p<0.001, \chi^2=37.42$) followed by pairwise comparisons using Wilcoxon's tests ($p<0.001, Z=19.50$ ours vs. no behavior, $p<0.001, Z=0.00$ ours vs. oracle). } 
	\label{fig:userstudy}
\end{figure}

\vspace{0.05in}
Participants were not told which condition they were seeing and 
as shown in Figure~\ref{fig:userstudy} participants rated the
behaviors of the players generated using our full behavior model as
significantly more realistic (e.g. more consistent with their
expectations of what the real-life player would do) than the baseline
configuration using no behavior model. The difference was particularly
pronounced when assessing recovery position ($\mu=3.76$ for ours
vs. $\mu=2.16$ for no behavior model), where real-life players have
well-known tendencies for where they recover.  The difference was less
pronounced for shot selection ($\mu=3.84$ for ours vs. $\mu=3.2$ for
no behavior model), likely because in real play,
players tend to vary their shot selection decisions (to keep an opponent off balance)
making it more challenging for participants to confidently judge any one specific shot selection
decision as atypical.

Finally, participants thought that the points reconstructed by our
oracle model generated the most realistic behaviors, suggesting that
while our behavior model is significantly better than the baseline
model there are still aspects of player behavior
that it does not yet capture.  Nevertheless, we note that many of the
participants also verbally told the experimenters that the task was
difficult, especially when considering the videos generated using
our behavior model or the oracle model. They often watched these
videos multiple times carefully looking for behavioral artifacts.

\subsection{Visual Quality of Player Sprites}

\begin{figure}[t]
	\centering
	\includegraphics[width=\columnwidth]{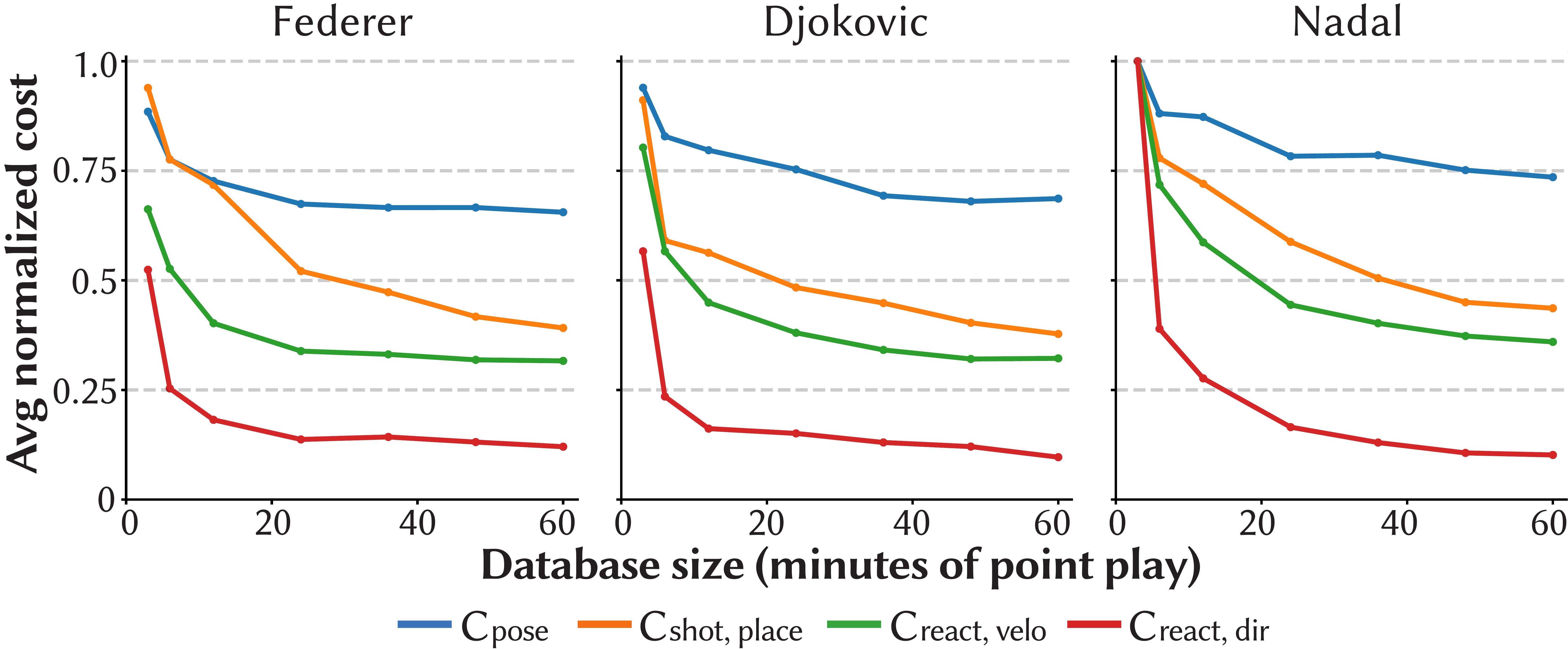}
	\caption{Clip search cost (averaged over 400 simulated rallies) decreases most quickly for Federer, suggesting that his controlled style of play yields a narrow distribution of player configurations. Except for $\mathbf{C_{react, dir}}$ (red line), Nadal's simulations have higher costs than those for Federer and Djokovic. The plots normalize each cost term by its maximum value encountered over all experiments (values across graphs are comparable).}
	\label{fig:database_size}
\end{figure}

\paragraph{Image quality}
Overall, our system generates visually realistic video sprites that often resemble broadcast video of Wimbledon matches.
However, visual artifacts do remain in generated output.
For example, player mattes exhibit artifacts due to errors in player segmentation,
in particular when the far-court player overlaps a linesperson on screen.
(Players and linespersons both wear white clothing.)
Second, appearance normalization is imperfect,
so transitions between clips with dramatically different lighting conditions
may exhibit appearance discontinuities.
Finally, 2D pose interpolation is insufficient to remove animation discontinuities
when transitions occur between frames with large pose differences.
While our system benefits from a viewer's gaze being directed at the opposing player during clip transitions
(they occur when the ball is on the other side of the court),
generating more realistic transitions is an interesting area of further work.

\paragraph{Sensitivity to clip database size}
To understand how clip search cost decreases with increasing amounts of source video content,
we measure the average value of components of the clip search cost metric (Sec.~\ref{sec:clip_cost}) as the database size is increased (Fig.~\ref{fig:database_size}). 
(We compute the average clip search cost over 400 simulated rallies for each player.)
Federer's results converge most quickly, suggesting that his iconic smooth, ``always well positioned'' playing style yields a narrower distribution of real-life configurations.  Similarly, Nadal's costs converge less quickly, and overall remain higher than Federer and Djokovic's, indicating greater diversity in his on-court play. 

\begin{figure}[t]
	\centering
	\includegraphics[width=\columnwidth]{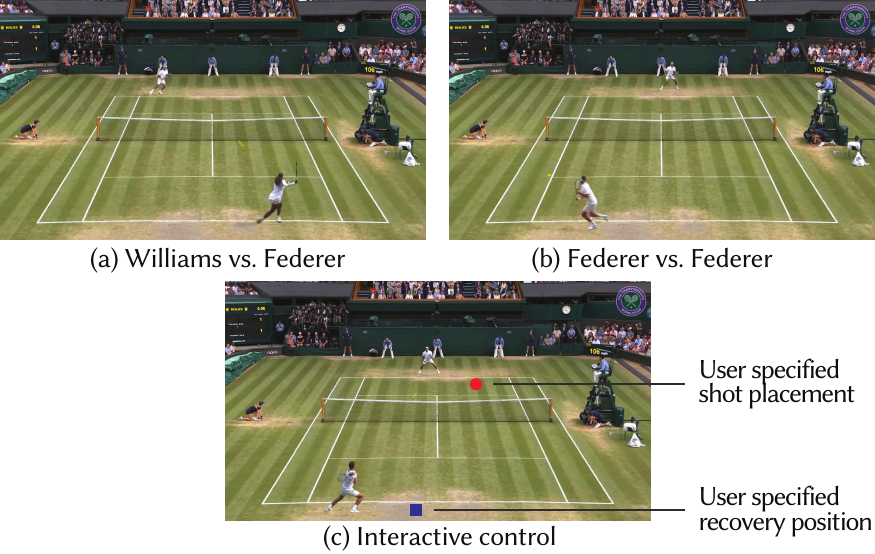}
	\caption{Our system can generate matchups that have not occurred in real life, such as (a) Roger Federer competing against Serena Williams, or (b) Federer competing against himself.  (c) A player's behavior goals can also be controlled by interactive user input.  We have also created an interface where the user clicks the court to set shot placement (red) and recovery position (blue) goals.} 
	\label{fig:applications}
\end{figure}

\subsection{Novel Matchups, Point Editing, and Interactive Control}
\label{sec:eval:interactive}

In addition to synthesizing points that depict tendencies observed in
real-life points between the two participating players,
we also use our system to generate novel matchups that have not occurred in real life.
For example, Fig.~\ref{fig:applications}-a,b depicts Roger Federer playing points against Serena Williams
and against himself on Wimbledon's Centre Court.
Since opponent-specific behavior models do not exist for these never-seen-before matchups,
we construct general behavior models for Federer and Williams by considering
all database clips where they are playing against right-handed opponents.  

Our system can also be used to edit or manipulate points that occurred in real life, such as modifying their outcome.
For example, the supplemental video contains an example where we modify a critical point from the 2019 Wimbledon final where Roger Federer misses a forehand shot that would have won him the championship.
(Federer famously ended up losing the match.)
Instead of making an error, the modified point depicts Federer hitting a forehand winner.

Finally, our system also supports interactive user control of players.
Fig.~\ref{fig:applications}-c illustrates one control mode implemented in our system, where
the near side player is controlled by the user and the far side player is controlled by the player behavior model.
In this demo, the user clicks on the court to indicate desired shot placement position (red dot) and player recovery position (blue square).  The near side player will attempt to meet these behavior goals in the next shot cycle following the click.  We refer the reader to the supplemental video for a demonstration of this interface.

\section{Discussion}
\label{sec:discussion}

We have demonstrated that realistic and controllable tennis player video sprites can be generated by analyzing the contents of single-viewpoint broadcast tennis footage.
By incorporating models of shot-cycle granularity player behavior into our notion of ``realism'',
our system generates videos where recognizable structure in rallies between professional tennis players emerges,
and players place shots and move around the court in a manner representative of real point play.
We believe this capability, combined with the interactive user control features of our system,
could enable new experiences in sports entertainment, visualization, and coaching.

Our database only contains a few thousand shots,
limiting the sophistication of the player behavior models we could consider.
It would be exciting to expand our video database to a much larger collection of tennis matches,
allowing us to consider use of more advanced behavior models,
such as those developed by the sports analytics community\,\cite{Wei:2016:shotforecasting,Wei:2016:thinedge}.
In general we believe there are interesting synergies between the desire to model human behavior in the graphics and sports analytics communities. 

We choose to work with single-viewpoint video because it is widely available,
and can be captured at sporting events of all levels.
However, most modern top-tier professional sports arenas now contain high-resolution, multi-camera rigs that accurately track player and ball trajectories for analytics or automated officiating\,\cite{StatsPerformSportVU,SecondSpectrum,Owens:2003:hawkEye}.
It will be interesting to consider how these richer input sources could improve the quality of our results.

Finally, our work makes extensive use of domain knowledge of tennis to generate realistic results.
This includes the shot cycle state machine to structure point synthesis,
the choice of shot selection and player court positioning outputs of player behavior models,
and the choice of input features provided to these behavior models.
It is likely that other racket sports may require different behavior model inputs
that better reflect nuances of decision-making in those sports,
however we believe that structure of the shot cycle state machine,
as well as our current behavior model outputs,
follow general racket sports principles and should pertain to sports such as 
table tennis, badminton, or squash as well.
It will be interesting to explore extensions of our system that apply these
structures to generate realistic player sprites for these sports.

\bibliographystyle{ACM-Reference-Format}
\bibliography{racket}


\end{document}